\documentstyle[prl,aps,twocolumn,epsfig]{revtex} 
\begin{document} 
\preprint{} 
\draft 

\title{Quantum Depletion of an Excited Condensate} 

\author{ Jacek Dziarmaga$^{1,2}$,
         Zbyszek P. Karkuszewski$^{1,2}$,
         and Krzysztof Sacha$^2$} 

\address{ 
${}^1$ Los Alamos National Laboratory, 
       Theory Division T-6, Los Alamos, 
       New Mexico 87545, USA \\ 
${}^2$ Intytut Fizyki Uniwersytetu Jagiello\'nskiego, 
       Reymonta 4, 30-059 Krak\'ow, Poland } 

\maketitle 

We analyze greying of the dark soliton in a Bose-Einstein condensate in
the limit of weak interaction between atoms. The condensate initially
prepared in the excited dark soliton state is loosing atoms because of
spontaneous quantum depletion. These atoms are depleted from the soliton
state into single particle states with nonzero density in the notch of the
soliton. As a result the image of the soliton is losing contrast. This
quantum depletion mechanism is efficient even at zero temperature when a
thermal cloud is absent.

\begin{abstract}

\end{abstract} 

\section{Introduction}
  Since the experimental creation of atomic Bose-Einstein condensates (BEC)
\cite{BEC}, BEC is a subject of intensive experimental and theoretical research.  The Gross-Pitaevskii
equation (GPE) \cite{GPE} is commonly used to describe with remarkable precision the ground
state properties and the dynamics of BEC. The GPE assumes that all atoms are in the same
single particle wave-function. In this sense it is a time-dependent mean field equation. This
equation is not an accurate description of the ground state in optical lattices. The
experiment in the optical lattice trap \cite{Orzel} demonstrates substantial squeezing of
the atomic ground state. As predicted in Ref.\cite{DDO}, squeezing of the atomic state
can also be achieved with a continuous phase contrast imaging of a condensate. The imaging
is a non-destructive weak measurement of the density of atoms which makes the phase of the
condensate wave-function ill defined. In this paper we ask a simple
question: is the GPE an accurate description of a collectively excited
condensate in a standard harmonic trap?

  The recent experiment by the Hannover group \cite{Hannover} strongly suggests a negative
answer. In Ref.\cite{Hannover} experimental results are carefully compared with simulations
of the GPE. The observed greying of the dark soliton is not consistent with the GPE. 
In this paper we analyze the problem in a weak particle interaction limit. 
The analysis allows us to relate greying of the dark soliton to a quantum depletion 
of atoms from the soliton state into single particle states with nonzero density in the 
soliton notch. This mechanism does not require any help from external agents like e.g. 
collisions with atoms from the thermal cloud \cite{Shlyapnikov}. The quantum depletion 
is present even at zero temperature. The condensate prepared in the excited dark soliton 
state has an intrinsic instability. 
Atoms coherently evolve away from the soliton condensate without help from any external
dissipation.

\section{ Gross-Pitaevskii equation (GPE)} 
$N$ bosonic atoms with repulsive s-wave two-body
interactions in a trap potential $V(x)$ are described by the $N$-body Schr\"odinger equation

\begin{eqnarray}
i\frac{\partial}{\partial t}\Psi&=&
\sum_{k=1}^N 
\left[
-\frac12 {\nabla}_k^2+
V({\bf x}_k)
\right] \Psi+a \sum_{k\neq l}^N \delta({\bf x}_k-{\bf x}_l)\; \Psi\;.
\label{SchE}
\end{eqnarray}
Here $\Psi(t,{\bf x_1},\dots,{\bf x_N})$ is a symmetric $N$-body wave-function. We use a
dimensionless form of the Schr\"odinger equation. When $N$ atoms are condensed in a single
particle state $\Phi(t,{\bf x})$ i.e.

\begin{equation}
\Psi(t,{\bf x}_1,\dots,{\bf x}_N)=
\prod_{k=1}^N
\Phi(t,{\bf x}_k)\;,
\label{product}
\end{equation}
then evolution of $\Phi(t,{\bf x})$ can be described by the GPE \cite{GPE}

\begin{equation}
i\frac{\partial}{\partial t}\Phi=
\left[
-\frac12{\bf \nabla}^2+     
V({\bf x})+
g\;\Phi^*\Phi
\right] \Phi 
\label{GPE}
\end{equation}
with $g\equiv(N-1)a$.
The derivation of Eq.(\ref{GPE}) from Eqs.(\ref{SchE},\ref{product}) assumes that the evolving
state of $N$ atoms remains in the subspace of product states (\ref{product}). In this paper we
show that this assumption is not satisfied for excited states of the GPE. Product states
(\ref{product}) are not eigenstates of exchange interactions which are neglected in the
derivation of the GPE. 

\subsection{ Symmetries of the GPE} 
GPE has symmetries similar to those of the one-body Schr\"odinger
equation. For example, in one dimension an (anti-)symmetric $\Phi(t,x)$,
$\Phi(t,-x)=(-)\Phi(t,x)$, remains (anti-)symmetric in the course of the time evolution.
In particular, a static dark soliton \cite{solit1}

\begin{equation}
\Phi(x)\;\sim\;
\tanh(\alpha x)
\label{darksoliton}
\end{equation}
in a symmetric trap remains antisymmetric and the density of atoms $n(x)=N|\Phi(x)|^2$ has a
zero at $x=0$. An antisymmetric dark soliton preserves its empty notch according to GPE.

\section{ Three mode approximation for small g} 
We assume an effectively one dimensional cigar-shaped harmonic trap i.e. one with
$\omega_{\bot}\gg\omega_x$ when the transversal motion of atoms is frozen in the transversal
ground state. The remaining $x$-dimension has the harmonic potential
$V(x)=\omega_x^2x^2/2$. For small $g$ we may truncate the single particle 
Hilbert space to the three lowest energy states (we choose oscillator units 
where $\omega_x=1$)

\begin{eqnarray}
u_0(x)&=&
\frac{ e^{-x^2/2} }{ \pi^{1/4} }\;,
\nonumber\\
u_1(x)&=&
x \frac{ 2^{1/2} e^{-x^2/2} }{ \pi^{1/4} } \;,
\nonumber\\
u_2(x)&=&
\left(x^2-\frac12\right)
\frac{ 2^{1/2} e^{-x^2/2} }{ \pi^{1/4} } \;.
\label{mody}
\end{eqnarray}
$u_0$ and $u_2$ are symmetric and $u_1$ is antisymmetric. 
The (symmetric) ground state of the GPE is

\begin{equation}
\Phi_0(x)\;=\;A\;u_0(x)\;+\;(1-A^2)^{1/2}\; u_2(x)\;,
\label{Phi0}
\end{equation}
with $A\in [0,1]$. $A=1$ for $g=0$ and decreases with increasing $g$. 
The admixture of $u_2(x)$ for nonzero $g$ is the closest that we can get to the Thomas-Fermi
limit in our 3-mode approximation valid for small $g$.

  The antisymmetric excited state of the GPE

\begin{equation}
\Phi_{\rm sol}(x)=u_1(x)
\label{Phisol}
\end{equation}
is a (small $g$) 3-mode analog of the dark soliton in Eq.(\ref{darksoliton}).

  To study quantum evolution we will use the second quantized Hamiltonian

\begin{equation}
\hat{H}=
\int dx\;
\left[
\frac12
\partial_x \hat{\Phi}^{\dagger}
\partial_x \hat{\Phi} +
V(x)\hat{\Phi}^{\dagger}\hat{\Phi}+
\frac{a_x}{2}
\hat{\Phi}^{\dagger}\hat{\Phi}^{\dagger}
\hat{\Phi}\hat{\Phi}
\right]\;.
\end{equation}
In the 3-mode approximation $\hat{\Phi}(x)=\hat{a}_0u_0(x)+\hat{a}_1u_1(x)+\hat{a}_2u_2(x)$,
where $\hat{a}_i$ are usual bosonic annihilation operators,
and the Hamiltonian becomes

\begin{eqnarray}
\hat{H}_3&=&
\frac12\hat{a}_0^{\dagger}\hat{a}_0+
\frac32\hat{a}_1^{\dagger}\hat{a}_1+
\frac52\hat{a}_2^{\dagger}\hat{a}_2+
\nonumber\\
&+&
I_{0000}\hat{a}_0^{\dagger}\hat{a}_0^{\dagger}\hat{a}_0\hat{a}_0+
I_{1111}\hat{a}_1^{\dagger}\hat{a}_1^{\dagger}\hat{a}_1\hat{a}_1+
I_{2222}\hat{a}_2^{\dagger}\hat{a}_2^{\dagger}\hat{a}_2\hat{a}_2+
\nonumber\\
&+&
I_{0011}
\left[
4\hat{a}_0^{\dagger}\hat{a}_1^{\dagger}\hat{a}_0\hat{a}_1+
\left(
\hat{a}_0^{\dagger}\hat{a}_0^{\dagger}\hat{a}_1\hat{a}_1
+{\rm h.c.}
\right)
\right]+
\nonumber\\
&+&
I_{0022}
\left[
4\hat{a}_0^{\dagger}\hat{a}_2^{\dagger}\hat{a}_0\hat{a}_2+
\left(
\hat{a}_0^{\dagger}\hat{a}_0^{\dagger}\hat{a}_2\hat{a}_2
+{\rm h.c.}
\right)
\right]+
\nonumber\\   
&+&
I_{1122}
\left[
4\hat{a}_1^{\dagger}\hat{a}_2^{\dagger}\hat{a}_1\hat{a}_2+
\left(
\hat{a}_2^{\dagger}\hat{a}_2^{\dagger}\hat{a}_1\hat{a}_1
+{\rm h.c.}
\right)
\right]+
\nonumber\\
&+&
I_{0112}
\left[
4\hat{a}_0^{\dagger}\hat{a}_1^{\dagger}\hat{a}_1\hat{a}_2+
2\hat{a}_0^{\dagger}\hat{a}_2^{\dagger}\hat{a}_1\hat{a}_1+
{\rm h.c.}
\right]+
\nonumber\\
&+&
I_{0002}
\left[
2\hat{a}_0^{\dagger}\hat{a}_0^{\dagger}\hat{a}_0\hat{a}_2+
{\rm h.c.}
\right]+
\nonumber\\
&+&
I_{0222}
\left[
2\hat{a}_0^{\dagger}\hat{a}_2^{\dagger}\hat{a}_2\hat{a}_2+
{\rm h.c.}
\right]
\label{H3}
\end{eqnarray}
with integrals defined by

\begin{equation}
I_{abcd}=\frac{a_x}{2}\int_{-\infty}^{+\infty} dx\; u_a(x)u_b(x)u_c(x)u_d(x) \;.
\end{equation}

\section{ Depletion of the ground state}

For noninteracting ($g=0$) particle case modes (\ref{mody}) constitute exact
eigenstates of the system. If $a_x\ne 0$ there are various couplings between
modes. The truncation of the Hilbert space to 3-modes only 
is a valid approximation if the interaction energy per atom does not shift 
significantly energy of an atom, i.e. $(N-1)I_{0000}<\omega_x$. That 
allows us to employ values of $g=(N-1)a_x$ up to 10. 

We have diagonalized
$\hat{H}_3$ for $N\leq 50$ and for $0\leq (N-1)a_x \leq 5$ in the Fock basis

\begin{equation}
|N_0,N_1,N_2\rangle=
\frac{ (\hat{a}^{\dagger}_0)^{N_0} 
       (\hat{a}^{\dagger}_1)^{N_1}
       (\hat{a}^{\dagger}_2)^{N_2} }
      {\sqrt{N_0!N_1!N_2!}}|0\rangle 
\end{equation}
with $N_0+N_1+N_2=N$. 
We find that, except for $(N-1)a_x=0$, the $N$-body ground state $|GS\rangle$ is not exactly a
product state like in Eqs.(\ref{Phi0},\ref{product}). The second quantized version of that
product state is

\begin{equation}
|A\rangle=
\frac{ \left[ A\hat{a}_0^{\dagger} +
              (1-A^2)^{1/2}\hat{a}_2^{\dagger} \right]^N }
      {\sqrt{N!}}|0\rangle\equiv
\frac{(\hat{a}^{\dagger}_A)^N}{\sqrt{N!}}|0\rangle\;.
\label{A}
\end{equation}
We find the product state which is the closest to the ground state $|GS\rangle$ with the help of
the number of atoms in a given product state $|A\rangle$ defined by $N_A=\langle
GS|\hat{a}^{\dagger}_A\hat{a}_A|GS\rangle$. The optimal product state $|A\rangle$ minimizes the
fraction of atoms depleted from the condensate

\begin{equation}
D_{\rm GS} \;=\; {\rm Min} \left( 1 \;-\; \frac{N_A}{N} \right)\;.
\end{equation}
The dependence of $D_{\rm GS}$ 
on the interaction strength $g=(N-1)a_x$ is shown in Fig.1.  We find
that $A$ decreases with increasing $(N-1)a_x$ (the condensate is transferred to the mode $u_2$)
and $D_{\rm GS}$ increases with $(N-1)a_x$ (increasing fraction of atoms is 
depleted from the condensate). 
As expected, see e.g. Ref.~\cite{castin}, for the ground state of the system, 
the depletion is a small effect.

\section{Depletion from the first excited state}

According to GPE the dark soliton
in our 3-mode approximation is the state (\ref{Phisol},\ref{product}), or
in the second quantized language 
\begin{equation}
|0,N,0\rangle.
\end{equation} 
The state
$|0,N,0\rangle$ is a stationary state of the GPE but it is {\it not} an
eigenstate of $\hat{H}_3$.  The quantum Hamiltonian is allowed to mix
antisymmetric modes with symmetric ones provided the symmetry with respect
to the parity operation in the full $N$-particle Hilbert space is
preserved. Various terms in $\hat{H}_3$, like e.g.
$\hat{a}_0^{\dagger}\hat{a}_2^{\dagger}\hat{a}_1\hat{a}_1$,
$\hat{a}_2^{\dagger}\hat{a}_2^{\dagger}\hat{a}_1\hat{a}_1$, or
$\hat{a}_0^{\dagger}\hat{a}_0^{\dagger}\hat{a}_1\hat{a}_1$, are mixing the
antisymmetric mode $1$ with symmetric modes $0$ and $2$. In more physical
terms, the vertex
$\hat{a}_0^{\dagger}\hat{a}_2^{\dagger}\hat{a}_1\hat{a}_1$ describes a
collision of two atoms in the antisymmetric soliton mode $1$ and their
transfer into two symmetric modes $0$ and $2$. The term
$\hat{a}_2^{\dagger}\hat{a}_2^{\dagger}\hat{a}_1\hat{a}_1$ is again a
collision of two atoms in the mode $1$ but this time both atoms are
transferred to the symmetric mode $2$. A similar vertex
$\hat{a}_0^{\dagger}\hat{a}_0^{\dagger}\hat{a}_1\hat{a}_1$ transfers pairs
of atoms from the mode $1$ to the symmetric mode $0$.

\subsection{Degenerate subspace}

Among these processes the most efficient is 
\begin{equation}
\hat{a}_0^{\dagger}\hat{a}_2^{\dagger}\hat{a}_1\hat{a}_1+{\rm h.c.},
\label{cruc}
\end{equation}
because such a transfer preserves single particle energy. For $a_x=0$, the
solitonic state $|0,N,0\rangle$ belongs to a degenerate subspace spanned
by vectors 
\begin{equation}
|i,N-2i,i\rangle. 
\end{equation}
When $a_x\neq 0$, the process (\ref{cruc})
couples these states among each other. Due to the degeneracy, an arbitrary
small $a_x$ is enough to mix $|0,N,0\rangle$ with other states from the
subspace. Even for $a_x\rightarrow 0$ all eigenstates of $\hat{H}_3$ in
the subspace have nonzero depletion --- even a very week interaction
between atoms induces a significant depletion of the soliton. For the
system in the ground state there is no similar process connecting
degenerate states and the depletion in the ground state is very small.

  We have calculated the depletion of the eigenstates, $|\phi\rangle$, 
of $\hat{H}_3$ from the soliton wave-function

\begin{equation}
D_S=1-\frac{N_S}{N}, 
\end{equation}
where $N_S=\langle\phi | \hat{a}_1^\dagger\hat a_1 |\phi\rangle$, for
$N=30$. In Fig.2 we show the depletion of the eigenstate which has the
smallest $D_S$ as a function of $(N-1)a_x$.  When $(N-1)a_x\rightarrow 0$,
the depletion does not tend to zero. This is because of the degeneracy of
the subspace the soliton state belongs to for $a_x=0$. Indeed, the
$|0,N,0\rangle$ state is not a limit of any eigenstates of $\hat{H}_3$
when $a_x$ decreases. Such an effect is obviously possible for a harmonic
potential only. For other traps, strong mixing of the soliton with the
$|i,N-2i,i\rangle$ states begins at finite $a_x$.  The insert of Fig.2
compares the density profiles of the GPE state $|0,30,0\rangle$ and of the
least depleted $\hat{H}_3$-eigenstate. The eigenstate has a substantial
nonzero density at $x=0$.

\subsection{Evolution from zero depletion state}

  In contrast to the GPE, in the full quantum evolution there is no
symmetry constraint to keep the notch of the initial soliton empty.  As
$|0,N,0\rangle$ is not an eigenstate of $\hat{H}_3$, the state of the
system $|\psi(t)\rangle$ prepared in the dark soliton state,
$|\psi(0)\rangle=|0,N,0\rangle$, evolves away from this state. As a result
of depletion the notch of the dark soliton is filling up and the soliton
is greying. In Fig.3 we show the contrast $C(t)$ as a function of time.
The contrast is defined by 

\begin{equation} 
C(t)=\frac{n_{\rm max}(t)-n(t,x=0)}{n_{\rm max}(t)+n(t,x=0)},
\end{equation} 
where $n_{\rm max}(t)$ is the maximal value of the density $n(t,x)$ at
time $t$. The insert of Fig.3 shows
the density $n(x)$ of the state (initially prepared as the dark soliton)
at different moments of time.  Due to the depletion the notch of the dark
soliton is filling up. Imporantly, depletion of the soliton reaches much greater 
value than the depletion of the corresponding least depleted eigenstate of the system.

  The time-scale on which the soliton contrast decays can be estimated
from below by $NI_{0000}$: the interaction energy of atoms in the ground
mode $u_0$. $I_{0000}\geq I_{abcd}$ so the interaction energy per atom in
the $N$-atom ground state can be used as an upper estimate for the
depletion time from the soliton condensate.  Employing this crude
approximation to the experimental situation \cite{Hannover} (i.e.
$N=1.5\cdot 10^5$ atoms of $^{87}$Rb in the trap with
$\omega_{\bot}=2\pi\times 425\,{\rm Hz}$ and $\omega_x=2\pi\times 14\,{\rm
Hz}$) we get for the minimal depletion time $t_{\rm
min}=2\pi/\Omega=0.6\,{\rm ms}$. The interaction energy per atom,
$\Omega=11.2\,{\rm kHz}$, has been estimated employing the Thomas-Fermi
approximation for a condensate wave-function. The minimal time $t_{\rm
min}$, is about 15 times shorter than the observed experimentally greying
time, i.e. $\sim 10$~ms \cite{Hannover}. We emphasize that this result is
meant to give a scale on which the evolution of a soliton is affected by
quantum depletion rather than a quantitative prediction for the
experiment. More accurate estimate for the Thomas-Fermi limit of large $g$
would have to include a large number of even and odd modes. In particular
it would have to take into account depopulation of even modes by
two-particle scattering to odd modes. Such calculations could in
principle be done in the spirit of the formalism of Ref.\cite{GZV}.
However, in Ref.\cite{DS} two of us present large $g$ calculations in the
framework of the Bogoliubov theory. They predict that a dark soliton will
loose contrast after around $10~{\rm ms}$.

\section{Conclusion} We have analyzed quantum depletion of the solitonic solution 
of the GPE for weak interaction between atoms [small $g$, see Eq.~(\ref{GPE})]. We have shown 
that a condensate prepared initially in the soliton state loses its contrast 
according to the quantum multiparticle time evolution.
Our 3-mode model is a suggestive example that the 
GPE cannot be used to describe large collective excitations of the
condensate. The model is a proper description
of the problem for a weak interaction between atoms only but it ilustrates essential 
processes responsible for the depletion of the solitonic wave-function. 

  It was suggested in Ref.\cite{Shlyapnikov,Hannover} that the deviations
from GPE can be possibly explained by dissipation due to collisions of the
condensate with the thermal cloud of non-condensed atoms. There is no doubt
that dissipation influences dynamics of the soliton in that experiment.
However, our analysis implies that a dark soliton may grey even in the absence 
of any thermal cloud. The notch may fill up with atoms quantum depleted from 
a condensate.

  Our analysis of the greying of the dark soliton may convey a wrong 
impression that a vortex in BEC \cite{dalibard} should also 
grey as a result of quantum depletion. Solitons are experimentally 
excited by means of the so-called phase imprinting method \cite{Hannover}. 
In our (small $g$) 3-mode approximation we model the result of such an excitation 
as the product state of the antisymmetric modes Eqs.~(\ref{Phisol},\ref{product}). 
In the case of a vortex 
state the situation is different. Indeed, vortices are created in a rotating 
(slightly asymmetric) cylindrical trap after achieving BEC or by cooling 
down the gas below the condensation temperature in an already rotating trap. 
Switching to the frame rotating with the trap one finds a single quantized 
vortex as the $N$-atom ground state that is an eigenstate of the 
$N$-atom Hamiltonian and as such it does not evolve in time. 
Eigenstates of a rotating condensate that consists of small number
of atoms has been analyzed recently in Ref.~\cite{ahsan}. It was observed that
indeed the ground state with a vortex reveals a noticeable depletion.
This ground state depletion is stationary, it does not grow in time.
However, vortices created by the phase imprinting method or subject
to suddenly stopped rotation are expected to reveal much larger quantum
depletion which may appear as greying.

{\it Note added:} Quantum depletion in a stationary solitonic state
has been predicted recently in \cite{law}. 

{\bf Acknowledgements.} The authors thank Diego Dalvit and Eddy Timmermans
for helpful discussions. J.D. and Z.K. were supported in part by NSA. K.S.
acknowledges support of KBN under project 5~P03B~088~21.

\begin{figure}
\centering
\caption{ The fraction of atoms $D_{GS}$ depleted from the condensate in 
the $N$-body
ground state as a function of the interaction strength $(N-1)a_x$ for several
values of $N=10, 30, 50$ (from top to bottom). }
\end{figure}

\begin{figure}
\centering
\caption{ The depletion from the soliton wave-function $|0,30,0\rangle$ 
of the least depleted eigenstate of $\hat{H}_3$ as a function of $(N-1)a_x$.
The insert
shows the density of atoms $n(x)$ in the dark soliton $|0,30,0\rangle$ 
(solid line)
and in the eigenstate (dashed line) at $(N-1)a_x=4.7$. }
\end{figure}

\begin{figure}
\centering    
\caption{ The contrast $C(t)$ of the evolving soliton for $(N-1)a_x=1$ 
and $N=20$. The contrast will go through revivals for times later than 40 due
to finite Hilbert space of the considered three mode approximation.
The insert shows densities $n(x)$ of the soliton at $t=10$ (solid line) and at $t=30$ (dashed line). }
\end{figure}

\end{document}